\documentclass[prl,twocolumn,showpacs,letterpaper,showpacs,superscriptaddress]{revtex4}
\usepackage{graphicx,amsmath,amssymb,amsfonts,latexsym,color,dcolumn,bm}

\begin{document}

\title{Probing quantum vacuum geometrical effects with cold atoms}

\author{Diego A. R. Dalvit}
\affiliation{Theoretical Division, Los Alamos National Laboratory, Los Alamos, NM 87545, USA}

\author{Paulo A. Maia Neto}
\affiliation{Instituto de F\'{\i}sica, UFRJ, CP 68528,   Rio de Janeiro,  RJ, 21941-972, Brazil}

\author{Astrid Lambrecht}
\author{Serge Reynaud}
\affiliation{Laboratoire Kastler Brossel, case 74,
CNRS, ENS, UPMC, Campus Jussieu, F-75252 Paris Cedex 05, France}

\date{\today}

\begin{abstract}
The lateral Casimir-Polder force between an atom and a corrugated surface 
should allow one to study experimentally non trivial geometrical effects
in quantum vacuum. Here, we derive the theoretical expression of this force 
in a scattering approach that accounts for the optical properties of the 
corrugated surface. We show that large corrections to the ``proximity force 
approximation'' could be measured using present-day technology with a 
Bose-Einstein condensate used as a vacuum field sensor. 
\end{abstract}

\pacs{42.50.Ct, 34.50.Dy, 03.75.Kk}

\maketitle

The Casimir force between bulk surfaces as well as the Casimir-Polder 
force between an atom and a surface \cite{CasimirPolder1948} are due to the 
reshaping of vacuum field fluctuations imposed by the boundary conditions \cite{Milonnibook}. 
In recent years there has been considerable progress in measurements of these forces
\cite{Cexperiments,CPexperiments}, opening the way for various applications 
in nanotechnology and quantum science \cite{applications}.

Most studies of the effect of geometry on the Casimir force are based on the proximity 
force approximation (PFA) \cite{Derjaguin1957}, which essentially amounts to an averaging over plane geometries. 
The related pairwise summation approach was used to calculate roughness corrections to
the force between an atom and a plate \cite{Galina2000}.
It is clear however that Casimir forces are 
not additive, except in the special case of very dilute media. Predictions beyond the
PFA have recently been obtained, for the normal component of the Casimir force between 
non-planar surfaces \cite{nonPFAnormal}, the effect of roughness \cite{EPL2005}, and
the lateral Casimir force between corrugated surfaces \cite{nonPFAlateral}. 

In this Letter we calculate non-trivial effects of geometry, \textit{i.e.} effects beyond the PFA,
on the Casimir-Polder force between an atom in its ground state and a corrugated surface. 
In contrast to the case of the lateral Casimir force between corrugated surfaces
\cite{Mohideen2002}, where the effect of reshaping vacuum field fluctuations is averaged over the surfaces, 
an atom is a local probe of the lateral Casimir-Polder force.
We show that deviations from the PFA can thus be much larger than for the force between
two surfaces. We first consider the simple case of sinusoidal corrugations and come
then to a periodically grooved surface where the deviation can be even more impressive.
We present estimations of the order of magnitude of the lateral Casimir-Polder force 
for a Bose-Einstein condensate trapped close to a corrugated surface \cite{Cornell,Schmiedmayer}.
These estimations indicate that an experimental demonstration of non trivial effects of geometry
on quantum vacuum should be within reach.

Our calculations will be based upon the scattering approach 
\cite{NJP2006} 
that has been developed to study the Casimir force between two surfaces in presence of stochastic roughness \cite{EPL2005} 
or corrugation \cite{nonPFAlateral}. This approach is used in a perturbative expansion with respect
to the corrugation amplitude $a$, which should be the smallest length-scale in the problem
$a \ll z_A, \lambda_c, \lambda_A,\lambda_0$, but allows one to deal with arbitrary relative
values of the atom/surface distance $z_A$, the corrugation wavelength $\lambda_c$  
and the typical wavelengths $\lambda_A$ and $\lambda_0$ characterizing the optical responses of atom and surface respectively. 
It will thus be possible to recover the trivial PFA limit 
$z_A, \lambda_A, \lambda_0 \ll \lambda_c$ as well as to explore non-trivial (beyond-PFA) effects, 
both for the van der Waals ($z_A \ll \lambda_A$) and Casimir-Polder ($\lambda_A \ll z_A$) regimes.

{\it Sinusoidal corrugation ---}
We first consider the case of an atom located above a sinusoidally corrugated surface.
The position components of the atom are denoted $x_A, y_A, z_A$ 
and the uni-axial corrugation described by a profile function $h(x)=h_0 \cos(k_cx)$,
measured from the plane $z=0$ (amplitude $h_0$ and wavelength $\lambda_c=2\pi/k_c$). 
In an expansion in powers of $h$, the scattering upon the corrugated surface
is given by zeroth-order specular reflection amplitudes and first-order non specular amplitudes. 
Higher-order terms will not be considered in this letter.

The Casimir-Polder energy of the atom above the surface is then written as
\begin{equation}
U_\mathrm{CP} = U_\mathrm{CP} ^{(0)}(z_A) + U_\mathrm{CP} ^{(1)}(z_A,x_A) .
\label{UCP}
\end{equation}
The zeroth-order term $U_\mathrm{CP} ^{(0)}$ 
is the standard Casimir-Polder potential between the atom and a plane plate, 
it depends only on specular reflection amplitudes. 
The first-order correction $U_\mathrm{CP} ^{(1)}$ depends on $x_A$ and thus gives rise
to the lateral Casimir-Polder force. It can be written in terms of
the non specular reflection amplitudes by using the techniques developed
in \cite{EPL2005,nonPFAlateral}. 

For the simple sinusoidal case, we find
\begin{equation}
U_\mathrm{CP} ^{(1)} = h_0 \cos(k_cx_A)g(k_c,z_A) .
\label{U1}
\end{equation}
The 
response function 
$g$ can be expressed in terms of 
the dynamic atomic polarizability $\alpha$ of the atom
(whose ground state is assumed to be spherically symmetric)
and of the first-order nonspecular amplitudes 
(same notation $R_{pp'}$ as in \cite{EPL2005})
\begin{eqnarray}
&&g({\bf k}, z_A) =  \frac{\hbar}{c^2 \epsilon_0} \int_0^{\infty} \frac{d\xi}{2\pi} \;  \xi^2 \alpha(i \xi)
\int \frac{d^2{\bf k}'}{(2\pi)^2} a_{{\bf k}', {\bf k}'-{\bf k}} ,
\nonumber\\
&&a_{{\bf k}', {\bf k}''} = \sum_{p',p''} 
\mbox{\boldmath\({\hat \epsilon}\)}_{p'}^{+}\cdot
\mbox{\boldmath\({\hat \epsilon}\)}_{p''}^{-}  \frac{e^{-(\kappa'+\kappa'')z_A}}{2\kappa''} 
R_{p'p''}({\bf k}', {\bf k}'') .
\label{g}
\end{eqnarray}
The bold letters ${\bf k}$ represent lateral wavevectors of the fields,
$p$ their polarizations (TE for transverse electric and TM for transverse magnetic). 
The roundtrip propagation of the field between the surface and the atom
is contained in the factor $\exp(-(\kappa'+\kappa'') z_A)$, with 
$\kappa=\sqrt{\xi^2/c^2 + k^2}$.
$\xi$ is the imaginary frequency of the field and $\epsilon_0$ the vacuum permittivity.

{\it Proximity Force Approximation ---}
The proximity force approximation (PFA) corresponds to the limiting case 
where the corrugation is very smooth with respect to the other length scales. 
We may thus replace in (\ref{U1}) the response function $g(k_c,z_A)$ by its 
limit $k_c\rightarrow0$ which can be shown to satisfy the ``proximity force theorem"
\begin{equation}
g(k_c\rightarrow0,z_A) = - \frac{\mathrm{d} U_\mathrm{CP} ^{(0)}(z_A)}{\mathrm{d}z_A} .
\label{gPFA}
\end{equation}
This identity holds for any specific optical response of the atom and bulk surface.
It is simply a consequence of the fact that $k_c \rightarrow 0$ corresponds 
to the specular limit of the nonspecular reflection amplitudes \cite{EPL2005}.

It follows from (\ref{gPFA}) that the Casimir-Polder interaction (\ref{UCP}) between the atom
and the corrugated surface can be obtained as 
$U_\mathrm{CP} \approx U_\mathrm{CP} ^{(0)} (z_A-h)$
where $U_\mathrm{CP} ^{(0)}$ is the ordinary Casimir-Polder potential
and $(z_A-h)$ the local separation distance between the atom and the surface.
This is essentially the ``proximity force approximation'' which is valid
when $g(k_c,z_A)$ may be replaced by $g(0,z_A)$. 
But it also follows from the preceding discussion that deviations from the
PFA prediction should be noticeable as soon as $g(k_c,z_A)$ differs from $g(0,z_A)$.
The deviations can be quantified by the ratio $\rho = g(k_c, z_A)/g(0, z_A)$.

We will see below that $\rho$ is smaller than unity for a sinusoidal corrugation,
which implies that the PFA overestimates the magnitude of the lateral effect in this case. Then, we will consider different kinds of corrugations
where this conclusion may be spectacularly modified.

{\it Perfect and real materials ---}
Assuming first that the corrugated surface can be described as perfectly reflecting, 
we deduce the nonspecular amplitudes from \cite{EPL2005} and
derive an explicit expression for the response function $g$. 
We write the results for the non-retarded (van der Waals) and 
the retarded (Casimir-Polder) regimes respectively 
\begin{eqnarray}
g(k_c, z_A) &=& - \frac{\hbar G(k_c z_A)}{64 \pi^2 \epsilon_0 z_A^4} \int_0^{\infty} d\xi \; \alpha(i \xi) 
\,,\quad z_A \ll \lambda_A \nonumber\\
G({\cal Z}) &=& {\cal Z}^2 [ 2 K_2({\cal Z}) + {\cal Z} K_3({\cal Z})] ;
\label{vdWperfect} \\
g(k_c, z_A) &=& - \frac{3 \hbar c \alpha(0)}{8 \pi^2  z_A^5} F(k_c z_A) 
\,,\quad z_A \gg \lambda_A \nonumber\\
F({\cal Z}) &=& e^{-{\cal Z}} (1+{\cal Z}+ 16 {\cal Z}^2/45 + {\cal Z}^3/45) .
\label{CPperfect}
\end{eqnarray} 
$K_2$ and $K_3$ are the modified Bessel functions of second and third order. 
As is usual, the Casimir-Polder expression depends only on the static 
polarizability $\alpha(0)$ of the atom. Note that the description of the surface as 
perfectly reflecting is more adapted to the retarded regime than to the non retarded one.
The PFA result corresponds to the limit ${\cal Z} \rightarrow 0$ in 
expressions (\ref{vdWperfect},\ref{CPperfect}).  

In Fig.~1 we plot the lateral potential $U^{(1)}$ as a function of $k_c z_A$ 
for a rubidium atom in front of a perfectly reflecting surface.
The dynamical polarizability data is obtained from \cite{Babb1999}
and the corrugation wavelength has been chosen as $\lambda_c=10\,\mu{\rm m}$.
The inset shows the deviation of the scattering result from the PFA. 
To give a number on deviation from the PFA, let us consider a separation $z_A= 2\mu$m, 
well within the CP regime, and a corrugation wavelength $\lambda_c=3.5\mu$m 
($k_c z_A \approx 3.55$), with $\rho\approx30 \%$, which means that the PFA
largely overestimates the magnitude of the lateral effect. 

\begin{figure}[t]
\includegraphics[width=8cm]{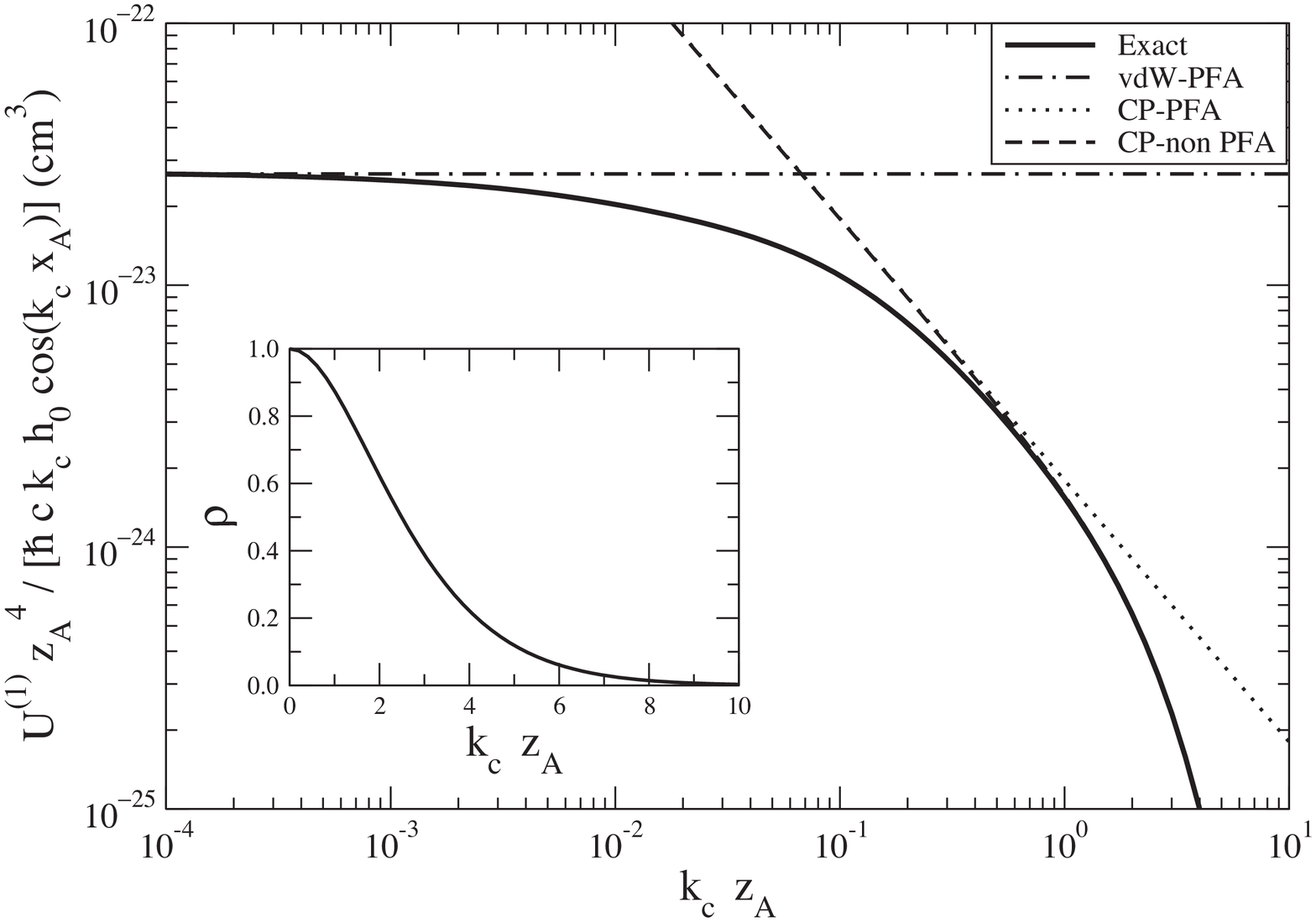}
\caption{Lateral potential energy $U^{(1)}$ for a rubidium atom in front of a 
perfectly reflecting surface with a sinusoidal corrugation of wavelength $\lambda_c = 10 \mu$m.
The inset shows the ratio $\rho$ that measures the deviation from the PFA.}
\end{figure}

It is important to consider the case of real materials. 
The lateral Casimir-Polder energy Eq.~(\ref{U1}) can thus be
evaluated by using the expression of the nonspecular reflection amplitudes $R_{pp'}$ 
in terms of the electric permittivity of the bulk material \cite{EPL2005}. 
The  frequency-dependent permittivity
along the imaginary frequency axis is evaluated by means of the Kramers-Kronig relations
in terms of the tabulated optical data for the different materials \cite{Palik}. 
As for the perfectly conducting surface, it turns out that PFA overestimates
the lateral Casimir-Polder force in the case of a sinusoidal corrugation, 
and that the function $\rho$ decreases 
exponentially to zero as $k_c z_A$ grows. This is due to the fact that 
the reflected field modes thus correspond to an
exponentially small propagation factor $\approx \exp(-k_{\rm c} z_A)$ in (\ref{g}). 
Specific results obtained for some materials of interest will be shown below.

\textit{BEC above a grooved surface ---}
We come now to the discussion of a configuration which should allow one
to show experimental non trivial effects of geometry on quantum vacuum, 
using the novel possibilities offered by cold atoms techniques.
A Bose-Einstein condensate (BEC) trapped in close proximity to a surface
has already been successfully used to observe the normal component of 
the Casimir-Polder force by measuring the frequency shift of the 
dipolar oscillation of the center-of-mass of the BEC \cite{Cornell}. 
For the lateral force we have in mind a setup where the long axis of an elongated BEC 
would be parallel to the corrugations,
while the lateral (along the $x$ direction) dipolar oscillation of the center-of-mass 
would be monitored as a function of time (see Fig.~2).
An interesting corrugation profile corresponds to periodical grooves.
If the atom is located above one plateau, the PFA predicts that the lateral 
Casimir-Polder force should vanish, since the energy is thus unchanged in a small lateral displacement.
A non vanishing force appearing when the atom is moved above the plateau
thus clearly signals a deviation from the PFA.
Note that, in an alternative experimental scheme, one could measure density variations along 
the $x$ direction as has been done in Ref.~\cite{Schmiedmayer}. 
In this case, the elongated BEC should preferably be aligned along the $x$-direction, 
and a density modulation along this direction above the plateau would be a signature of a
nontrivial geometry effect. 

\begin{figure}[t]
\includegraphics[width=5cm]{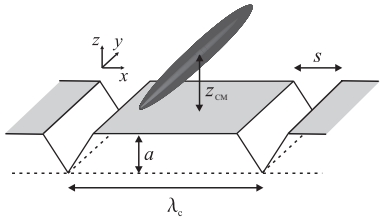}
\caption{ An elongated Bose-Einstein condensate is trapped 
close to a grooved surface, near the plateau region.
Its  dipole oscillation frequency
along the $x$ direction is shifted due to the lateral force effect.}
\end{figure}

For the periodical grooves of Fig.~2, the potential (\ref{U1}) 
has to be replaced by a sum over
Fourier components $a_n$ of the corrugation profile 
(assumed to be even for simplicity)
$h(x)=\displaystyle{\sum_n} a_n \cos(nk_cx)$
\begin{equation}
U_\mathrm{CP} ^{(1)} = \sum_{n=0}^{\infty} a_n \cos(nk_cx_A) g(nk_c,z_A) .
\label{U1sum}
\end{equation}
The PFA is now recovered when the response function $g(nk_c,z_A)$ may be replaced
by the limiting expression $g(0,z_A)$ for all values of $n$ significantly contributing to the profile.
When this is the case, $U_\mathrm{CP} ^{(1)}$ has the same profile as the corrugation itself. 
Otherwise, as $z_A$ increases
multiplication by $g(nk_c,z_A)$ renders  the contribution of higher orders 
comparatively smaller.
When $k_cz_A$ is much larger than unity in particular, the exponential decrease already discussed 
for $g$ implies that the first order $n=1$  dominates the sum (\ref{U1sum}) (apart from the 
irrelevant $n=0$ term) and that
the potential is approximately  sinusoidal.

{\it Frequency shifts for the dipolar oscillation ---}
In order to discuss the feasability of the experiment, we first evaluate
the relative frequency shift
$\gamma_0=(\omega_{{\rm CP},x} - \omega_x)/\omega_x$
 for a single rubidium atom of mass $m$ 
located at a distance $z_A$ from the plateau shown  in Fig.~2.
Here $\omega_x$ is the trap frequency along the $x$ direction and 
$\omega_{{\rm CP}, x}$ is the oscillation frequency in the presence
of the Casimir-Polder force. 
In Fig. 3 we plot the relative frequency shifts 
as functions of $k_c z_A$ for $z_A=2\mu$m,   groove width $s=\lambda_c/2,$
and depth $a=250$nm. 
We consider different materials for the bulk surface, namely a good conductor (gold), 
a semiconductor (silicon), and a dielectric (fused silica).

Assuming PFA, the shift would
vanish since the potential is locally flat 
on top of the plateau in this case. Fig.~3 shows that $\gamma_0$ is indeed very small for $k_c z_A<1.$ But 
as $k_c z_A$ increases, $\gamma_0$ develops a peak and then decays exponentially for $k_c z_A\gg 1.$
The maximal frequency shift decreases
as the atom-surface separation grows, reaching values lower than $10^{-5}$ for distances
greater than $z_A =3\mu$m.
Given the reported sensitivity of $10^{-5}-10^{-4}$ for relative frequency
shift measurements \cite{Cornell}, we expect that the effect could be measured for distances
below $3\mu$m both for gold and silicon surfaces, while being on the border of detectability 
for fused silica. 
Different materials induce a variety of error sources, including stray electric fields 
(especially from spurious charges in insulating surfaces) and magnetic fields 
(from adsorbed atoms in conducting and semiconducting surfaces). 
Ultimately the optimal choice of material would come from a tradeoff
between maximizing the signal and minimizing the error sources \cite{Cornell}.

The single-atom case described up to now can be applied to a BEC of Thomas-Fermi radius $R$
only in the ``point-like" limit  $R \ll z_{\rm CM}, \lambda_c$,
with $z_A$ merely replaced by the center-of-mass coordinate $z_{\rm CM}$.
Otherwise, we have to calculate finite-size corrections.
To this aim, we consider an elongated rubidium BEC with trap frequencies 
$\omega_y \ll \omega_x = \omega_z$, and a two-dimensional density profile in the 
Thomas-Fermi limit given by $(15/6 \pi) \left(R^2-(x^2+z^2)\right)^{3/2}/R^5$.  
When averaging the single-atom frequency shift $\gamma_0$ over the
two-dimensional density, the resulting shift $\gamma$ is found to increase 
as a function of $R$ as shown in the inset of Fig.~3. 
The frequency shift due to the lateral Casimir-Polder force on a finite-size BEC should thus be detectable for distances
below $3 \mu$m and a radius of, say,  $R \approx 1 \mu$m. 
Note that non-linear corrections due to the finite amplitude $\delta_x$ of oscillations along the $x$ direction
can be estimated along the lines of \cite{Antezza2004}. For the lateral Casimir-Polder
force they are proportional to $k_c^2 \delta_x^2/8$, and for typical values  
$\delta_x=0.5 \mu$m and $\lambda_c=4 \mu$m they induce only a small 
decrease of the relative frequency shift of about $8\%$.

\begin{figure}[t]
\includegraphics[width=8cm]{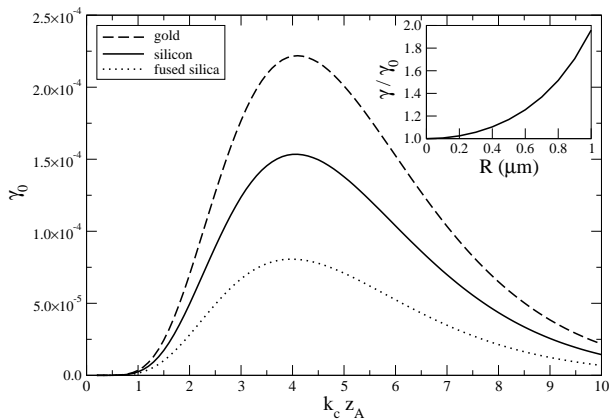}
\caption{Relative frequency shift due to the lateral Casimir-Polder
force for a single rubidium atom located at a distance $z_A=2 \mu$m from 
the corrugated surface in Fig.~2 made of different materials. 
The unperturbed frequency  is $\omega_x/2 \pi = 229$ Hz, the groove
width is $s=\lambda_c/2$, and the depth $a=250$ nm.
The inset shows the corrections to the relative frequency shift for a finite size BEC of Thomas-Fermi
radius $R$ oscillating at a distance $z_{\rm CM}=2 \mu$m from a silicon grooved surface of period $\lambda_c=4 \mu$m. For gold and fused silica
we obtain very similar results, which are indistinguishable
on the scale of the inset.}
\end{figure}

{\it Conclusion ---}  
Novel cold atoms techniques open a promising way of investigating nontrivial geometrical effects
on quantum vacuum. 
An interesting feature of atoms, with respect to macroscopic objects, is that they can be used as local probes 
of quantum vacuum fluctuations as they are reshaped when scattered by nontrivial boundaries. 
For example an atom or a BEC above the plateau of a grooved plate would see no lateral effect,
should the widely used proximity force approximation be exact. 
The results of the present letter, based upon a scattering approach and accounting for a realistic optical response
for both atoms and bulk material, show that the non-trivial beyond-PFA effect should 
be measurable with a BEC using available technology.

\acknowledgments

We are grateful to James Babb for providing us with the dynamic polarizability data for rubidium, and to
Mauro Antezza and Malcolm Boshier for interesting discussions.
P.A.M.N. thanks  CNPq, CAPES
and Institutos do Mil\^enio de Informa\c c\~ao Qu\^antica e Nanoci\^encias
for financial support. A.L. acknowledges partial financial support 
by the European Contract STRP 12142 NANOCASE.

\end{document}